\documentstyle[12pt]{article}

\newcommand{\be}{\begin{equation}}
\newcommand{\ee}{\end{equation}}
\newcommand{\bea}{\begin{eqnarray}}
\newcommand{\eea}{\end{eqnarray}}
\newcommand{\nn}{\nonumber}

\begin{document}
\baselineskip=15pt
\null\vskip -2cm
\hfill {YUMS 94-20} \\
\vspace{2mm}
\hfill {SNUTP 94-76}
\vskip 2.5cm
\begin{center}

{\Large Hawking Radiation of Dirac Fields \\ in the (2+1)-Dimensional
        Black Hole space-time.}\\
\vspace{3mm}

\vskip 2.0cm

{Seungjoon Hyun%
\footnote{email: {\sc hyun@phya.yonsei.ac.kr}} },
{Yong-Seon Song%
\footnote{email: {\sc song@phya.yonsei.ac.kr}} },
{and}
{Jae Hyung Yee%
\footnote{email: {\sc jhyee@phya.yonsei.ac.kr}} }

\vskip .5cm

{\sl Department of Physics\\
     and Institute for Mathematical Sciences\\
     Yonsei University\\
     Seoul 120-749, Korea}

\vskip 1.5cm

{\bf Abstract}
\end{center}
\vspace{.3cm}

\baselineskip=18pt

\noindent

By calculating the response function, we study the Hawking radiation of
massless Dirac fields in the (2+1)-dimensional black hole geometry. We
find that the response function has Planck distributions,
with the temperature that agrees with the previous results obtained for
the scalar field cases.
We also find the Green's functions in (2+1)-dimensional Einstein static
universe and anti de-Sitter space.
\vfill

\newpage

\baselineskip=18pt

\section{Introduction}
\indent
One of the fascinating aspects of (2+1)-dimensional quantum field
theories is the bose-fermi transmutation \cite{pol}. It appears when
the matter fields are coupled to the U(1) Chern-Simons gauge field. It
would be interesting to see whether the similar phenomena appear in
other theories and situations as well. Sometime ago Takagi has shown
\cite{Takagi} that in the Rindler background, the detector
observes the Hawking radiation with transmuted statistics depending on
the number of space-time dimensions.

Recently it has been observed \cite{LO} that the response function for
the scalar field in the (2+1)-dimensional black hole background
has Fermi-Dirac distribution, while the expectation value
of number operator, $<N>$, shows \cite{HLY} Planck distribution.

The underlying physical reason for these phenomena is not understood
yet, however, just like in the case of Rindler background, it may be
the universal features in the (2+1)-dimensional black hole physics.

In this paper we study the response function for the Dirac fields in
the (2+1)-dimensional black hole geometry and show that the distribution
is Planckian, and thus extend the statistics flip phenomena to the Dirac
fields.

We first compute find the Green's functions
in the Einstein static universe (ESU)
and anti de-Sitter space (AdS). We then we use the method of
images to calculate the Green's function in the black hole geometry,
by using the fact that the black hole space-time is given by identifying
points in AdS with space-like Killing vector.
This method has been used in \cite{LO} to calculate the Green's functions
in the (2+1)-dimensional black hole geometry.

The response function is given by the Fourier transformation of the
Green's function. The general form is too complicated to compute exactly,
hence we
restrict ourselves in the limiting case of large black hole mass M
with the detector in the asymptotic region.

This paper is organized as follows. In section 2 we briefly review
the geometry of (2+1)-dimensional black hole. In section 3 we calculate
the Green's functions in the ESU, AdS and black hole space-time.
In section 4 we calculate the response function and show
that it is Planckian. In section 5 we draw some conclusions.

\section{The Geometry of the (2+1)-dimensional Black Hole}
\indent

The Ban\~{a}dos-Teitelboim-Zanelli(BTZ) metric \cite{Ban} for
three-dimensional black hole solution is
\begin{equation}
ds^2=-N^2(r) dt^2+N^{-2}(r)dr^2+r^2(N^{\varphi}dt+d\varphi)^2,
\label{met}
\ee
where
\begin{eqnarray}
N^2&=&-M+\frac{r^2}{l^2}+\frac{J}{4r^2}, \nn \\
N_\varphi&=&-\frac{J}{2 r^2},
\label{lapse}
\end{eqnarray}
and $l$ is related to the cosmological constant by
$\lambda=-l^{-2}$.

We consider the case of $J=0$:
\be
ds^2=-(\frac{r^2}{l^2}-M) dt^2+(\frac{r^2}{l^2}-M)^{-1} dr^2
     +r^2 d\varphi^2,
\label{met1}
\ee
where $\varphi$ is identified with a period $2 \pi $.
This coordinate system covers the part of AdS which
can be realized as the three-dimensional hyperboloid,
\be
v^2-w^2 -x^2+y^2=-l^2,
\ee
in a four-dimensional space with metric
\be
ds^2=dv^2-dw^2-dx^2+dy^2 .
\ee
The parametrization for the black hole geometry (\ref{met1}) is given by
\begin{eqnarray}
v&=&\sqrt{\frac{r^2}{M}-l^2} \;\; \cosh\;{(\sqrt{M} t/l)}, \nn \\
w&=&\sqrt{\frac{r^2}{M}-l^2} \;\; \sinh\;{(\sqrt{M} t/l)}, \nn \\
x&=&\frac{r}{\sqrt{M}} \; \cosh \,{(\sqrt{M}\varphi)},   \nn \\
y&=&\frac{r}{\sqrt{M}} \; \sinh \,{(\sqrt{M}\varphi)},
\end{eqnarray}
with identification $\varphi \, \rightarrow \, \varphi+2\pi n$.

On the other hand, the parametrization which covers the whole AdS space
is given by
\begin{eqnarray}
v=l \;\tan{\,{\chi}} \;\cos{\,{\theta}},                    \nn \\
w=l \;\sin{\,{\tau}} \;\sec{\,{\chi}},                      \nn \\
x=l \;\cos{\,{\tau}} \;\sec{\,{\chi}},                      \nn \\
y=l \;\tan{\,{\chi}} \;\sin{\,{\theta}},
\end{eqnarray}
where $0 \leq \chi \leq \pi/2, \;\;
 0 \leq  \theta < 2 \pi$ and $0 \leq \tau  \leq 2 \pi$.
The metric can be read as
\be
ds^2=l^2\; \sec^2{\chi}[-d\tau^2+d \chi^2+\sin^2{\chi}\; d \theta^2],
\label{kr}
\ee
which is conformal to the half of the ESU with the metric:
\be
ds_E^2=-d\tau^2+d\chi^2+\sin^2{\chi}\; d \theta^2,
\label{ESU}
\ee
where $0 \leq \chi \leq \pi$.
These metrics are related by  ${g_{\mu\nu}}=\Omega^2 g_{\mu\nu}^E$
where $\Omega$ , the conformal factor, is given by $\Omega=l\sec{\chi}$.

AdS is well-known to have not only a closed time-like curve,
but also a time-like boundary at spatial infinity
through which information can propagate.

To avoid the closed time-like curves, we consider the universal covering
space of AdS (CAdS) in which $\tau$ takes any real value \cite{Ellis}.

One way to resolve the problem of time-like boundary is to
follow the prescription of \cite{Avis}, i.e, to take the Cauchy
surface as a pair of surfaces
$\{\tau,\;0 \leq \chi \leq \pi/2 \} $ and $\{ \tau+\pi,\;
 0 \leq \chi \leq \pi/2  \} $
in CAdS, which are causally disconnected. It has been used in \cite{HLY}
to study the Hawking radiation of scalar field.
In this paper we follow another prescription in which appropriate
boundary conditions at infinity, $\chi=\frac{\pi}{2}$, are taken.
We will explicitly construct the Green's functions in AdS and black hole
space-time which satisfy either Dirichlet, $\Psi(\chi=\frac{\pi}{2})=0$,
or Neumann, $\frac{\partial}{\partial\chi}\Psi(\chi=\frac{\pi}{2})=0$,
boundary condition.

\section{Green's functions in the (2+1)-dimensional black hole}
\indent

In this section we derive Green's functions of the massless two component
Dirac field $\Psi$ in (2+1)-dimensional black hole space-time, which
satisfies the Dirac equation,
\be
\gamma^{\mu}\bigtriangledown_{\mu}\Psi
=\gamma^a e_a^{\mu}\bigtriangledown_{\mu}\Psi = 0,
\ee
where\\
\be
\bigtriangledown_{\mu}  =  \partial_{\mu}+\frac{1}{2}\sigma_{ab}
\omega_{\mu}^{ab},\\
\ee
and $e^a_{\mu}$ and $\omega^{ab}_{\mu}$ denote the dreibein and spin
connection, respectively.
We choose our $\gamma$-matrices as ;\\
\be
\gamma_0  =  i\sigma_3,\hspace{10mm}\gamma_1  =
\sigma_1,\hspace{10mm}\gamma_2  =  -\sigma_2, \\
\ee
and\\
\be
\sigma_{ab}  =  \frac{1}{4}[\gamma_{a},\gamma_{b}]  =  \frac{1}{2}
\varepsilon_{abc}\gamma^{c}.
\ee
In order to calculate the Green's functions in black hole space-time,
we start from the fermions in ESU.
The Dirac eq.(10) in the ESU reads
\be
-i\partial_{\tau}\Psi_{+}+(\partial_{x}+i\sin^{-1}{x}\partial_{\theta}
+\frac{1}{2}\cot{x})\Psi_{-}=0,
\ee
\be
(\partial_{x}-i\sin^{-1}{x}\partial_{\theta}+\frac{1}{2}\cot{x})\Psi_{-}
+i\partial_{\tau}\Psi_{-}=0,
\ee
for two-component Dirac field $\Psi^E={\Psi_{+}\choose\Psi_{-}}$.
The mode solutions for these coupled equations are, so-called,
spin-spherical harmonics (or generalized spherical harmonics), which
have been considered in several different contexts \cite{NP, GMS};
\be
\Psi^E=e^{-iw\tau}e^{im\theta} {-iu_{m,-\frac{1}{2}}^l (\cos{\chi})
\choose u_{m,\frac{1}{2}}^l (\cos{\chi})},
\ee
where $\omega=l+\frac{1}{2},$ $l\in Z+\frac{1}{2},$ $-l\leq m\leq l.$
These spin-spherical harmonics are related to the Jacobi polynomials
as ;
\be
u_{m,\pm\frac{1}{2}}(\mu)=(1-\mu)^{\frac{\alpha}{2}}
(1+\mu)^{\frac{\beta}{2}}P^{\alpha,\beta}_{s}(\mu),
\ee
where
\be
\alpha=\mid\pm\frac{1}{2}-m\mid,\hspace{1cm}
\beta=\mid\pm\frac{1}{2}+m\mid,\hspace{1cm}
s=l-\frac{1}{2}(\alpha+\beta).
\ee
The Jacobi polynomials satisfy the following orthogonality relations;
\bea
\lefteqn{\int^1_{-1}(1-\mu)^{\alpha}(1+\mu)^{\beta}P^{(\alpha, \beta)}_n(\mu)
P^{(\alpha, \beta)}_m(\mu)d\mu} \nn\\
&&=\frac{2^{\alpha+\beta}}{2n+\alpha+\beta+1}
\frac{\Gamma(n+\alpha+1)\Gamma(n+\beta+1)}{n!\Gamma(n+\alpha+\beta+1)}
\delta_{n, m},
\eea
which ensure the orthogonality of $u^l_{m, \pm\frac{1}{2}}$.

One may understand this as follows.
The spatial part of ESU is 2-sphere, $S^2$. Killing vectors in this $S^2$
are nothing but the generators of angular momentum SU(2). For the
spin-half fields, we need to consider the total angular momentum and then
two components correspond to the components with $m\pm\frac{1}{2}$.
In this way one can easily generalize to higher spin fields in
(2+1)-dimensional ESU. For example, the mode solutions for spin-1 fields
are given by vector spherical harmonics.

The relevant two point Green's function in ESU is
\begin{eqnarray}
G_{E}(x,x^{'})&=&\langle0\mid\Psi^E\stackrel{-}{\Psi^E}\mid0\rangle \nn\\
&=&\sum_{l,m}(\Psi_{l,m})(\stackrel{-}{\Psi}_{l,m})
={G_{(1,1)}\hspace{0.7cm}G_{(1,2)}\choose
G_{(2,1)}\hspace{0.7cm}G_{(2,2)}} \nn\\
&=&\sum_{l,m}ie^{-iw(\tau-\tau^{'})}e^{im(\theta-\theta^{'})}
{u_{m,-\frac{1}{2}}u^*_{m,-\frac{1}{2}}\hspace{1cm}
iu_{m,-\frac{1}{2}}u^*_{m,\frac{1}{2}}\choose
iu_{m,-\frac{1}{2}}u^*_{m,\frac{1}{2}}\hspace{1cm}
u_{m,\frac{1}{2}}u^*_{m,\frac{1}{2}}}.
\end{eqnarray}
Using $u^{*}_{m+s}(\cos{x})=(-1)^{m+s}u_{s,m}(\cos{x})$ and the generalized
addition theorem \cite{GMS},
\bea
\lefteqn{\sum^{l}_{m=-l}e^{im(\theta-\theta^{'})}u_{m,s_i}(\cos{\chi})
u^*_{m,s_j}(\cos{\chi^{'}})}\nn\\
&=&\sum^{l}_{m=-1}(-1)^{m+s_j}e^{im(\theta-\theta^{'})}u_{s_i,m}
(\cos{\chi^{'}})u_{m,s_j}(\cos{\chi})
\nn\\
&=&(-1)^{s_i}e^{-is_i\varphi_{1}}e^{-is_j\varphi_{2}}
(1+\cos{\gamma})^{\frac{1}{2}}P^{(0,s_i+s_j)}(\cos{\gamma}),
\eea
where
\be
\cos{\gamma}=\cos{\chi'}\cos{\chi}+\sin{\chi'}\sin{\chi}
\cos{(\theta-\theta')},
\ee
and
\be
e^{\frac{i}{2}(\varphi_1+\varphi_2)}=\frac{
[\sqrt{(1+\cos{\chi})(1+\cos{\chi'})}e^{i\frac{\theta-\theta'+\pi}{2}}-
\sqrt{(1-\cos{\chi})(1-\cos{\chi'})}e^{-i\frac{\theta-\theta'+\pi}{2}}]}
{\sqrt{2(1+\cos{\gamma})}},
\ee
\be
e^{\frac{i}{2}(\varphi_2-\varphi_1)}=\frac{
[\sqrt{(1+\cos{\chi})(1-\cos{\chi'})}e^{i\frac{\theta-\theta'+\pi}{2}}+
\sqrt{(1-\cos{\chi})(1+\cos{\chi'})}e^{-i\frac{\theta-\theta'+\pi}{2}}]}
{\sqrt{2(1+\cos{\gamma})}},
\ee
we get
\bea
\lefteqn{G_{(i, j)}(x,x^{'})=Ke^{-\frac{i}{2}(\tau-\tau^{'})}
e^{i(s_i\varphi_1+s_j\varphi_2)}(1+\cos{\gamma})^{\frac{1}{2}}}\nn\\
&&\times\sum^{\infty}_{l=0}P_{l}^{(0,s_i+s_j)}(\cos{\gamma})
e^{-il(\tau-\tau^{'})}.
\eea
Then by using \cite{As},
\be
\sum^{\infty}_{l=0}P_{l}^{(0, s)}(\mu)z^{n}=
\frac{2^s}{R(1+z+R)^s},
\label{sum}
\ee
where $R=\sqrt{1-2\mu z+z^2}$,
we finally obtain
\bea
G_{(1,1)}&=&\frac{Ke^{\frac{i}{2}(\varphi_1+\varphi_2)}
(1+\cos{\gamma})^{\frac{1}{2}}}
{(\cos{\triangle\tau}-\cos{\gamma})^{\frac{1}{2}}
[\sqrt{2}\cos{\frac{\triangle\tau}{2}}+(\cos{\triangle\tau}
-\cos{\gamma})^\frac{1}{2}]},\nn\\
G_{(1,2)}&=&\frac{(-1)^\frac{3}{2}Ke^{\frac{1}{2}(\varphi_2-\varphi_1)}
(1-\cos{\gamma})^{\frac{1}{2}}}
{(\cos{\triangle\tau}-\cos{\gamma})^{\frac{1}{2}}
[-\sqrt{2}\cos{\frac{\triangle\tau}{2}}
+(\cos{\triangle\tau}-\cos{\gamma})^\frac{1}{2}]},\nn\\
G_{(2,1)}&=&-e^{i(\varphi_1-\varphi_2)}G_{(1,2)},\nn\\
G_{(2,2)}&=&-e^{-i(\varphi_1+\varphi_2)}G_{(1,1)}.
\eea
In eq.(\ref{sum}), which holds for $-1<\mu<1$ and $\mid z\mid<1,$
we give $\triangle\tau$ an infinitesimal negative imaginary part, i.e.
$\triangle\tau-i\varepsilon$, and the square root
$\sqrt{(\cos{\triangle\tau}-\cos{\gamma})}$
is defined with a branch cut along the negative real axis and the argument
in it is between $(-\pi, \pi)$.

We define another Green's function $\stackrel{\sim}{G}_E(x,x')$ by
\be
\stackrel{\sim}{G}_E(x,x')=G(\stackrel{\sim}{x},x'),
\ee
where $\stackrel{\sim}{x}=(\tau,\pi-\chi,\theta)$.
One can easily see that
\be
G^{\pm}_E(x, x')=G_E(x, x')\pm \stackrel{\sim}{G}_E(x, x')
\ee
satisfy Neumann $(+)$ and Dirichlet $(-)$ boundary conditions, respectively.

The Green's functions in AdS can be derived from the Green's functions in
ESU by the following conformal transformation;
\be
g_{\mu\nu}=\Omega^2g^E_{\mu\nu},\hspace{1cm}\Omega=l\sec{\chi}.
\ee
As the field $\Psi$ transforms under the conformal transformation as
\be
\Psi=\Omega^{-1}\Psi^E
=l^{-1}\cos{\chi}\Psi^E,
\ee
the Green's functions in AdS are given by
\be
G^{\pm}_A=\frac{1}{l^2}\cos{\chi}\cos{\chi'}G^{\pm}_E.
\ee
One can easily find the Green's functions in the black hole space-time,
which is constructed from (2+1)-dimensional AdS by identifying
$\varphi\rightarrow\varphi+2\pi n.$ By using the method of images the
Green's function $G_{BH}$ in black hole space-time can be written as
\be
G^{\pm}_{BH}(x, x')=\sum^{\infty}_{n=-\infty}G^{\pm}_A(x, x'_n),
\ee
where $x'=(t', r', \varphi')$ and $x'_n=(t', r', \varphi'+2\pi n).$

\section{The Response Function}
\indent
In this section we calculate the response of DeWitt point-like detector
\cite{Takagi, De}
coupled to the density of the Dirac field through the Lagrangian,
\be
L_{int}=M(\tau)\stackrel{-}{\psi}(x(\tau))\psi(x(\tau)),
\ee
where M is the monopole operator of the detector, and $\tau$ is the
detector's proper time. The transition rate of the detector for the field
from $E_1$ to $E_2$ is given by
\be
R(E_1/E_2)=\mid<E_2\mid M(0)\mid E_1>\mid^2 F_T(E_2-E_1),
\ee
where
\be
F_T(\omega)=\int^{\infty}_{-\infty}e^{-i\omega\tau}S(\tau)
d(\tau),
\ee
with $S(\tau-\tau')\equiv<0\mid\stackrel{-}{\psi}(x(\tau))\psi(x(\tau))
\stackrel{-}{\psi}(x(\tau'))\psi(x(\tau'))\mid 0>.$

Using Wick's theorem $S(\tau-\tau')$ can be expressed as,
\be
S(\tau-\tau')=tr[G(x(\tau), x(\tau'))G(x(\tau'), x(\tau))].
\ee

After performing the Fourier transformation, $F_T$ can be written as
\be
F_T(\omega)=\int^{\infty}_{-\infty}\frac{d\omega'}{2\pi}
F_{ij}(\omega-\omega')F_{ji}(\omega),
\ee
where the response function $F(\omega)$ is given by
\be
F_{ij}(\omega)=\int^{\infty}_{-\infty}
e^{-i\omega\triangle\tau}G_{(i,j)}(\triangle\tau)d(\triangle\tau),
\ee
and
\be
G_{(i,j)}(\triangle\tau)=G_{(i,j)}(x(\tau), x(\tau+\triangle\tau)).
\ee
We consider the two point functions $G^{+}$ which satisfy Neumann boundary
condition.
The detector's motion is chosen to be stationary in the black hole
coordinates, so that the world line of detector $x(\tau)$ is given by
$x(\tau)=(\tau/\sqrt{\frac{r^2}{l^2}-M},r,\phi).$
For simplicity we choose $x=(\frac{\tau}{\sqrt{\frac{r^2}{l^2}-M}}, \gamma,
0)$ and $x'=(0, \gamma, 2\pi n\sqrt{M}).$
This choice of $\phi$ and $t$ should not change the results since
$\frac{\partial}{\partial \phi}$ and $\frac{\partial}{\partial t}$ are
killing vector fields in the black hole geometry.
In terms of 4-dimensional coordinates, these correspond to
\bea
v&=&\sqrt{\frac{r^2}{M}-l^2}\cosh{\frac{\sqrt{M}t}{l}},\nn\\
w&=&\sqrt{\frac{r^2}{M}-l^2}\sinh{\frac{\sqrt{M}t}{l}},\nn\\
x&=&\frac{r}{\sqrt{M}},\nn\\
y&=&0,
\eea
and\\
\bea
v'&=&\sqrt{\frac{r^2}{M}-l^2},\nn\\
w'&=&0,\nn\\
x'&=&\frac{r}{\sqrt{M}}\cosh{2\pi n\sqrt{M}},\nn\\
y'&=&\frac{r}{\sqrt{M}}\sinh{2\pi n\sqrt{M}}.
\eea
In the asymptotic region, $r\rightarrow\infty$, the response function is
essentially reduced to the integrals
\begin{eqnarray}
& &\lefteqn{I_{(n)}(\omega)}\nn\\
&=&\int^{\infty}_{-\infty}dz\frac{e^{-i\frac{\omega}{2\pi T} z}}
{(\cosh{\alpha_{n}}-\cosh{z})^{\frac{1}{2}}
[(\cosh{\alpha_{n}})^{\frac{1}{2}}(1+\cosh{z})^{\frac{1}{2}}+
(\cosh{\alpha_n}-\cosh{z})^{\frac{1}{2}}]}\nn\\
&=&\int^{-\alpha_n}_{-\infty}dz\frac{e^{-i\frac{\omega}{2\pi T} z}}
{-i(\cosh{z}-\cosh{\alpha_n})^{\frac{1}{2}}
[(\cosh{\alpha_n})^{\frac{1}{2}}(1+\cosh{z})^{\frac{1}{2}}
-i(\cosh{z}-\cosh{\alpha_n})^{\frac{1}{2}}]}\nn\\
&+&\int^{-\alpha_n}_{\alpha_n}dz\frac{e^{-i\frac{\omega}{2\pi T} z}}
{(\cosh{\alpha_n}-\cosh{z})^{\frac{1}{2}}
[(\cosh{\alpha_n})^{\frac{1}{2}}(1+\cosh{z})^{\frac{1}{2}}
+(\cosh{\alpha_n}-\cosh{z})^{\frac{1}{2}}]}\nn\\
&+&\int^{\infty}_{\alpha_n}dz\frac{e^{-i\frac{\omega}{2\pi T} z}}
{i(\cosh{z}-\cosh{\alpha_n})^{\frac{1}{2}}
[(\cosh{\alpha_n})^{\frac{1}{2}}(1+\cosh{z})^{\frac{1}{2}}+
i(\cosh{z}-\cosh{\alpha_n})^{\frac{1}{2}}]}\nn\\
&=&I_1+I_2+I_3,
\end{eqnarray}
where $\alpha_n=2\pi n\sqrt{M}$, and the local temperature $T$ is given by
$T=\frac{1}{2\pi l\sqrt{\frac{r^2}{M}-l^2}}.$ This is related to the
temperature of the black hole $T_0=\frac{\sqrt{M}}{2\pi n}$ by the Tolman
relation $T=(g_{00})^{-\frac{1}{2}}T_0$ \cite{Tolman}.
In the case of $n=0$, the integral is given by
\begin{eqnarray}
I_{(0)}(\omega)
&=&\int^{\infty}_{-\infty}dz\frac{e^{-i\frac{\omega}{2\pi T} z}}
{-\cosh{z}-1+i\sinh{z}}\nn\\
&=&\frac{2\pi}{1-e^{\omega /T}},
\end{eqnarray}
from the contour integration.
For non-zero n, under the replacements
\be
t=e^z,\hspace{1cm}\beta_n=e^{\alpha_n},\hspace{1cm}\frac{t}{\beta_n}=y,
\ee
$I_3$ can be written as
\be
I_3=\int^{\infty}_{1} dy \frac{2y^{-i\frac{\omega}{2\pi T}}(\beta_n)^
{-i\frac{\omega}{2\pi T}-1}}
{i(y-1)^{\frac{1}{2}}(y-\frac{1}{\beta^2_n})^{\frac{1}{2}}
[\sqrt{\frac{\beta^n+\beta^{-1}_{n}}{2}}(y+\frac{1}{\beta_n})
+i(y-1)^{\frac{1}{2}}(y-\frac{1}{\beta^2_n})^{\frac{1}{2}}]}.
\ee
In the limit $\beta_n \rightarrow \infty$ (the large M limit), and change of
variables $x=\frac{1}{y}$, $I_3$ can be written as
\bea
\lefteqn{I_3\sim -2\sqrt{2} i\beta_n^{-\frac{3}{2}-i\frac{\omega}{2\pi T}}
\int^{1}_{0}dxx^{i\frac{\omega}{2\pi T}}(1-x)^{\frac{1}{2}}}\nn\\
&=&-2\sqrt{2} i\beta_n^{-\frac{3}{2}-i\frac{\omega}{2\pi T}}
B(1+i\frac{\omega}{2\pi T},\frac{1}{2})\nn\\
&=& 2\sqrt{2 \pi} \beta_n^{-\frac{3}{2}-i\frac{\omega}{2\pi T}}
\coth{\frac{\omega}{2T}}.
\eea
Similarly, we can calculate $I_1, I_2$ and the results are
\be
I_1\sim i2 \sqrt{2 \pi} \beta^{i\frac{\omega}{2\pi T}-\frac{3}{2}}_{n}
(i\coth{\frac{\omega}{2T}})\frac{\Gamma(-\frac{1}{2}+
i\frac{\omega}{2\pi T})}{\Gamma(i\frac{\omega}{2\pi T})},
\ee
\be 
I_2\sim 2\sqrt{2\pi}[\beta_n^{-\frac{3}{2}-i\frac{\omega}{2\pi T}}
\frac{\Gamma(-\frac{1}{2}-i\frac{\omega}{2\pi T})}
{\Gamma(-i\frac{\omega}{2\pi T})}
+\beta_n^{-\frac{3}{2}+i\frac{\omega}{2\pi T}}
\frac{\Gamma(-\frac{1}{2}+i\frac{\omega}{2\pi T})}
{\Gamma(i\frac{\omega}{2\pi T})}].
\ee
Therefore $I_{(n)}$ is given by
\be
I_{(n)}(\omega)=-\frac{4\sqrt{2\pi}}{e^{\frac{\omega}{T}}-1}
[\beta_n^{-\frac{3}{2}-i\frac{\omega}{2\pi T}}
\frac{\Gamma(-\frac{1}{2}-i\frac{\omega}{2\pi T})}
{\Gamma(i\frac{\omega}{2\pi T})}
+\beta_n^{-\frac{3}{2}+i\frac{\omega}{2\pi T}}
\frac{\Gamma(-\frac{1}{2}+i\frac{\omega}{2\pi T})}
{\Gamma(i\frac{\omega}{2\pi T})}],
\ee
and, after performing the infinite sum, the response function can
be written as
\be
F(\omega)=C\frac{M}{r^2} \frac{1}{e^{\frac{\omega}{T}}-1}
\{ -2\sqrt{2}\pi {1\\\ 0 \choose 0\\\ 1}
-4\sqrt{2\pi} {\ 1\\\ -1 \choose -1\\\\\ 1}
[\frac{\Gamma(-\frac{1}{2}-i\frac{\omega}{2\pi T})}
{\Gamma(i\frac{\omega}{2\pi T})}
+\frac{\Gamma(-\frac{1}{2}+i\frac{\omega}{2\pi T})}
{\Gamma(i\frac{\omega}{2\pi T})}]\}.
\ee
where C is a constant.
In the limit $T\rightarrow 0$, which is the case in the asymptotic region,
or in the limit $\omega\rightarrow\infty$, the response function becomes
\be
F(\omega)=C\frac{M}{r^2} \frac{1}{e^{\frac{\omega}{T}}-1}
[-2\sqrt{2}\pi {1\\\ 0 \choose 0\\\ 1}
-4\sqrt{2\pi} \sqrt{\frac{4\pi T}{\omega}}
{\ 1\\\ -1 \choose -1\\\\\ 1}].
\ee
This response function clearly shows that the distribution is thermal,
with the temperature $T$, and furthermore it is Planckian.

\section{Conclusion}
\indent
In this paper, we have studied the Hawking radiation of Dirac fields
in the (2+1)-dimensional black hole, by calculating the response function
of the DeWitt point-like detector.
By way of the calculations, we have also obtained the Green's functions
in ESU and AdS.
The Hawking temperature for Dirac fields agrees with the one obtained from
the scalar field case \cite{HLY}.

Most notably, we have found the Planck distribution for Dirac fields
in the (2+1)-dimensional black hole background.
This shows that statistics flip phenomenon of the response function is a
rather general feature of the Hawking radiation.
However, we should point out that
the expectation value of the number operator, $<N>$ does not show such
statistics flip as has been shown in ref. \cite{HLY},
in the case of scalar field.
It is not yet clear how we should interpret these results physically.

\newpage
\vspace{3cm}
\hspace{4cm}
{\bf Acknowledgments} \\
\vspace{0.7cm}

This work was supported in part by the Korean Science and
Engineering Foundation, Center for Theoretical Physics (S.N.U.) and
the Basic Science Research Institute Program, the Ministry of Education
Project NO. BSRI-94-2425.


\newpage

\begin{thebibliography}{99}
\bibitem{pol} A. M. Polyakov,
              Mod. Phys. Lett. {\bf A3} 455 (1988).
\bibitem{Takagi} S. Takagi,
              Theo. Phys. Suppl. {\bf 88} 1 (1986).
\bibitem{LO} G. Lifschytz and M. Ortiz,
              Phys. Rev. {\bf D49} 1929 (1994).
\bibitem{HLY} S. Hyun, G. H. Lee and J. H. Yee,
              Phys. Lett. {\bf B322} 182 (1994).
\bibitem{Ban} M. Ba\~{n}ados, C. Teitelboim, and J. Zanelli,
              Phys. Rev. Lett. {\bf 69} 1849 (1992).
\bibitem{Ellis} S. Hawking and G. F. R. Ellis,
                {\it The Large Scale Structure of Spacetime},
                (Cambridge University Press, Cambridge, 1973).
\bibitem{Avis} S. J. Avis, C. J. Isham and D. Storey,
              Phys. Rev. {\bf D18} 3565 (1978).
\bibitem{NP} E. T. Newman and R. Penrose,
              J. Math. Phys. {\bf 7} 863 (1966):\\
             J. N. Goldberg, A. J.Macfarlane, E. T. Newman, F. Rohlich and
             E. C. G. Sudarshan,
             {\it ibid} {\bf 8} 2155 (1967).
\bibitem{GMS} I. M. Gel'fand, R. A. Minlos and Z. Y. Shapiro,
             {\it Representations of the rotation and Lorentz groups and
             their applications,}
             (Pergamon press, Oxford, 1963).
\bibitem{As}  M. Abramowitz and I. A. Stegun,
             {\it Handbook of Mathematical Functions with Formulas, Graphs,
             and Mathematical Tables,}
             (National Bureau of Standards, 1963).
\bibitem{De} B. S. DeWitt, in {\it General Relativity}, ed. S. W. Hawking and
             W. Israel (Cambridge University Press, 1979).
\bibitem{Tolman} R. C. Tolman,
                {\it Relativity, Thermodynamics and Cosmology}
                (Oxford, UK, 1931).

\end{thebibliography}
\end{document}